# Detection of Moving Targets using Coherent Optical Frequency Domain Reflectometry

**Esther Renner[1], Max Koeppel[1], Bernhard Schmauss[1,2]**
[1]*Institute of Microwaves and Photonics, Friedrich-Alexander University Erlangen-Nürnberg (FAU), Cauerstr. 9, 91058 Erlangen, Germany.*
[2]*Max Planck Institute for the Science of Light, Staudtstr. 2, 91058 Erlangen, Germany.*
esther.renner@fau.de

**Abstract:** We present Doppler-corrected position and velocity measurement with a fiber-coupled COFDR system based on the FMCW radar principle for high precision localization applications. A high measurement accuracy and the ability to track targets are demonstrated. © 2020 The Author(s)

## 1. Introduction

Coherent optical frequency domain reflectometry (COFDR) is an interferometric localization method inspired by the frequency-modulated continuous-wave (FMCW) radar. It is well established in characterization and testing of fiber optical components and in distributed optical sensing [1]. Advantages of COFDR over other reflectometry techniques are high achievable resolution (µm-range), very high sensitivity and fast interrogation speed over intermediate measurement ranges [2].

Laser radar or lidar (light detection and ranging) systems are widely known for the measurement of range, velocity and vibration [3]. Especially coherent laser radar systems and FMCW-lidar systems are based on the same approach as frequency-swept COFDR [4,5]. For all mentioned techniques, a light source is frequency-modulated and the beam is split into two paths, a measurement path and a reference path to serve as local oscillator (LO). Mixing of the reflected signal from the target and the LO signal results in a beat signal at the receiver [6]. The frequency of the beat signal contains information regarding the distance of the target and, in case of a moving target, radial velocity.

In this paper, we present the results of Doppler-corrected position and velocity measurements of a moving target with an all fiber-coupled COFDR system. The presented COFDR system was used in an earlier stage for characterization in an additive manufacturing process [7] and for localization of 'flying' particles in a hollow-core photonic-crystal fiber [8]. The system is able to detect targets the size of a few micrometers with very high resolution and high repeatability. The Doppler-COFDR approach is distinguished by its specifications in range, resolution and target size from conventional lidar systems. Therefore, it opens up the possibility for new applications in additive manufacturing, test and measuring technology or particle tracking. Thus, it might be applicable for the detection of microparticle vibration and velocity while tracking the Doppler-corrected position over a length of a few meters. Here, we proof the suitability of the proposed COFDR system for velocity measurement as a basis for further research in high-precision localization and tracking applications.

## 2. Measurement setup and principle

The COFDR setup is based on a fiber-coupled Mach-Zehnder interferometer, as shown in Fig. 1. The *Luna Phoenix 1400* laser system was used as a main component, including a tunable laser source (TLS) and two photodetectors (PD). The TLS can emit a highly linear frequency ramp in the wavelength range of 1515 nm to 1565 nm, corresponding to a maximum frequency bandwidth of ~6 THz, with a sweep speed of 12.6 THz/s. In the first step, the beam of the TLS is split into two parts using a 94/6 polarization-maintaining (PM) beam-splitter. The minor part of the beam power (6 percent) is guided through an auxiliary-interferometer, whereas 94 percent of the beam power are directed towards the measurement-interferometer and thus the moving target. The measurement-interferometer is set up using standard single-mode (SM) fiber components. A 99/1 fiber-coupler is used to split the beam in measurement and reference path. In the measurement path (shown in green), a circulator guides the light to the free-space section. Subsequently, the light is coupled out of the fiber with a collimator and directed towards the moving target. The moving target was realized using a reflector mirror mounted on an electro-mechanical translation stage (brush motor controller *Thorlabs TDC001* and translation stage *Thorlabs MTS50/M-Z8*). The translation stage is able to move with a speed of up to 0.5 mm/s. The reflected light from the target then is coupled back into the fiber and the paths are recombined with a 50/50 beam coupler. A polarization controller is included in the reference path (shown in red) to maximize interference and a standard SM-fiber to compensate path length difference, i.e. defining a zero plane. The signal of the measurement-interferometer is detected at a photodetector (PD) of the laser system.

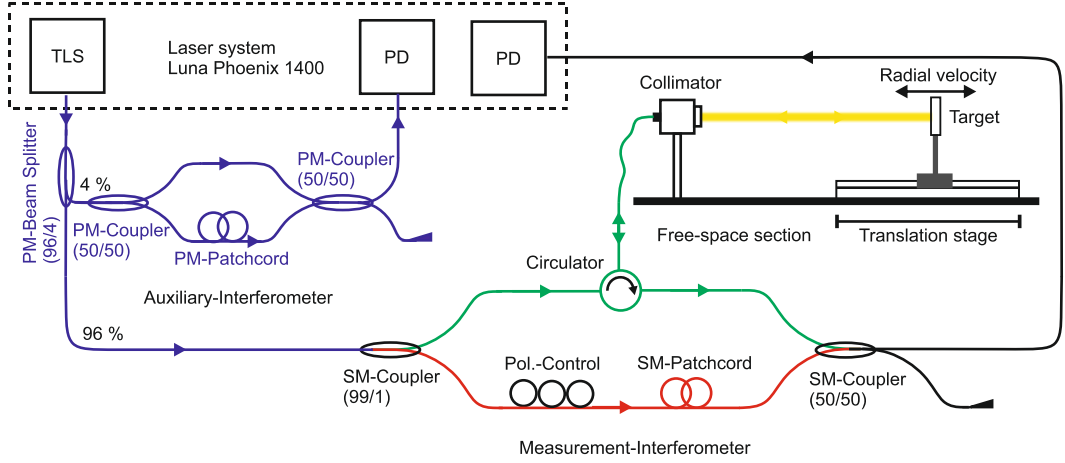

Fig. 1. Experimental setup of the fiber-coupled, frequency-swept COFDR system, including an auxiliary-interferometer and a free-space section for the integration of the moving target.

The fiber-coupled auxiliary-interferometer with a fixed path difference is employed using PM components. The observed beat-signal is detected at the second detector of the laser system and used to generate equally spaced frequency steps for the compensation of artefacts in the wavelength vector. The beat signals from both interferometers are acquired with the two internal PDs of the laser system and further processed on a computer. The beat signal is Fourier-transformed to obtain the beat frequency information.

For a stationary target this beat frequency depends on the propagation time difference $\tau = 2z_0/c$, where $z_0$ is the distance to the target, $c = c_0/n$ the speed of light and $n$ the refractive index of the medium. The beat frequency is related to the distance of the target by $f_b = \gamma_f \tau = 2\gamma_f z_0/c$ with frequency sweep rate $\gamma_f$ [7]. The time-dependent position of a linearly moving target with radial velocity $v$ is given by $z(t) = z_0 + vt$. Likewise, this results in a time-dependent propagation time difference:

$$\tau(t) = 2\frac{z(t)}{c} = \frac{2z_0}{c} + \frac{2v}{c}t. \tag{1}$$

The electrical field of the emitted laser beam incident on the measurement-interferometer can be written in complex notation as

$$E(t) = E_0 \exp\bigl(j(2\pi f_0 t + \pi \gamma_f t^2)\bigr), \tag{2}$$

where $E_0$ denotes the amplitude of the field and $f_0$ the start frequency of the sweep. This emitted field is separated in reference and measurement path. Thus, assuming ideal polarization and neglecting losses, the recombined electrical field incident on the detector can be expressed as

$$E_{\text{det}}(t) = E_{\text{ref}}(t) + E_{\text{meas}}(t) = \tfrac{1}{2}[(1-\kappa)E(t) + \kappa\sqrt{R}E(t-\tau(t))] \tag{3}$$

with target reflectivity $\sqrt{R}$ and constant $\kappa$ comprising coupling to the measurement path. The measured intensity $I(t)$ is proportional to the modulus squared of $E_{\text{det}}(t)$:

$$I(t) \propto |E_{\text{det}}(t)|^2. \tag{4}$$

By inserting equations (1) and (2) into (3), the intensity can be calculated following expression (4). Only the AC part of the intensity contains information on the propagation time difference. Therefore, the beat frequency is obtained by derivation of the phase of the cosine function contained in the intensity signal. Consequently, the beat frequency of a moving target is completely described by

$$f_{b,\text{mov}} = \underbrace{\frac{2\gamma_f z_0}{c}}_{\text{(I)}} + \underbrace{\frac{2f_0 v}{c}}_{\text{(II)}} - \underbrace{\frac{4\gamma_f z_0 v}{c^2}}_{\text{(III)}} + \underbrace{\left(\frac{4\gamma_f v}{c} - \frac{4\gamma_f v^2}{c^2}\right)t}_{\text{(IV)}}. \tag{5}$$

Term (I) of equation (5) contains the distance information and (II) the Doppler-frequency shift. Term (III) is a mixing product containing the distance distortion due to Doppler shift and (IV) is a chirp, comprising the movement during the sweep time. Terms (III) and (IV) can be neglected, if the target velocity is low compared to the speed of light and the sweep time short compared to the propagation time difference. Resulting, the beat frequency simplifies to $f_{b,mov} = f_z + f_D$. The distance information $f_z$ and Doppler shift $f_D$ are identified by processing two measurements with different sweep direction according to:

$$f_z = \tfrac{1}{2}(f_{b,up} - f_{b,down}) = \tfrac{2\gamma_f z_0}{c} \text{ and } f_D = \tfrac{1}{2}(f_{b,up} + f_{b,down}) = \tfrac{2f_0 v}{c}. \tag{6}$$

## 3. Experimental results

For the detection of a moving target, data are acquired during frequency up- and down-sweep of the laser source to determine the Doppler shift. The sweep bandwidth of the TLS was set to 1 nm from 1540 nm to 1541 nm and the sweep rate to 12.6 THz/s to perform fast sweeps and neglect the chirp. Fig. 2 shows the beat spectra obtained from the beat signal at the PD after Fourier transformation. The inset illustrates the influence of the Doppler shift on the beat frequency for up- and down-sweep. A fiber-end reflection at the collimator serves as a reference point. As the fiber-end reflection is stationary, it does not comprise a frequency shift due to the Doppler effect and therefore appears at the same beat frequency for up- and down-sweep. However, the reflection of the moving target shows up at different beat frequencies for up- and down-sweep because of the induced Doppler-frequency shift $f_D$. The Doppler shift is calculated from the two detected beat frequencies and the true position and the radial velocity of the target are estimated by equation (6).

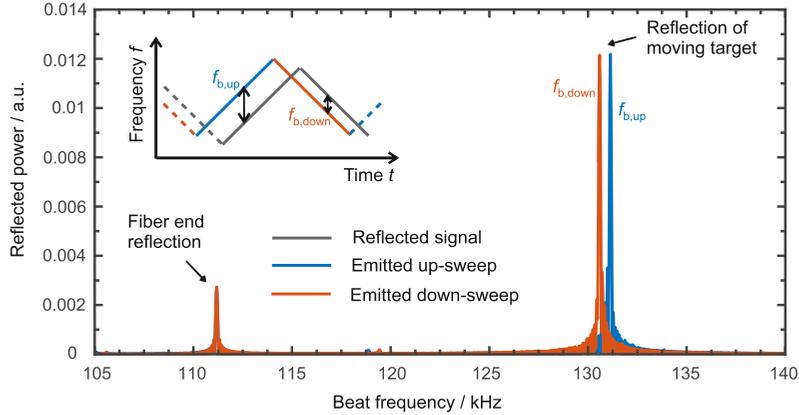

Fig. 2. Measured spectrum of the beat frequency for the moving target. Due to the movement of the target, the observed reflection is Doppler-shifted to $f_z \pm f_D$ for the frequency up- and down-sweep, respectively.

To determine the accuracy of the measured position and velocity, 10 measurements were taken with velocities from 0 mm/s up to 0.5 mm/s in steps of 0.05 mm/s. The reflector target starts moving from the zero position and each measurement is taken after 3 s to allow the motor controller to reach the specified velocity. This leads to a linear increase of the true position with velocity. The results for the position accuracy are compared to the read-out position from the translation stage as shown in Fig. 3 (a). The standard deviation (std.-dev.) is increasing with velocity for both measured and read-out position. The position accuracy of the translation stage is given to be 290 µm as specified by the vendor. The maximum observed std.-dev. for the measurement is 583 µm and therefore lies in the same order of magnitude as the read-out value. The mean error between read-out and measurement is 73 µm, which is lower than the accuracy of the translation stage. The results for the velocity accuracy are shown in Fig. 3 (b). As expected, the std.-dev. increases with faster target velocities. A maximum std.-dev. of 49 µm/s is observed. The accuracy of the velocity setting for the translation stage is specified by the vendor as 0.25 mm/s. Because of the low accuracy of the set velocity of the translation stage, the velocity measured by the COFDR system is assumed to be more accurate than the set velocity.

In a next step, a different translation stage was included in the setup to validate the system for faster moving targets. Fig. 3 (c) shows an example of target tracking over 80 s with a target moving with a radial velocity of approximately 20 mm/s and changing direction after 23 cm. This proofs that even tracking of faster targets is possible and that the direction of movement can be determined from the sign of the measured radial velocity.

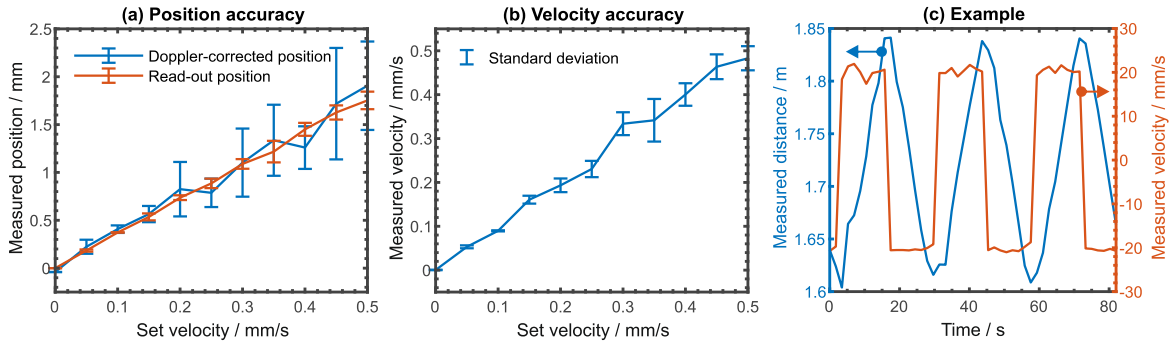

Fig. 3. Doppler-corrected results regarding positon (a) and velocity accuracy (b). In (c) an example for target position and velocity tracking with a target speed of 20 mm/s over 80 s was performed. The target was moved back and forth over a distance of 23 cm with change of direction, which leads to a change of the sign of the radial velocity.

## 4. Conclusion

We have demonstrated velocity measurement and position tracking for moving targets based on Doppler shift with a COFDR system for high precision localization. The position and velocity measurement with the presented COFDR system shows more accurate results than the read-out values of the translation stage. However, we plan to validate our results in further experiments with a new translation stage achieving higher accuracy for the set values. Nevertheless, a trade-off between two-point resolution and target speed has to be taken into account. Both are limited by the sweep speed of the TLS as the Doppler-dependency of the beat frequency leads to broadening of the reflection peak. Broadening increases with higher target velocities compared to the sweep speed. With an increase in sweep speed of the laser leading to a shorter measurement time, the COFDR approach opens up the possibility of an even more accurate determination of the Doppler shift, higher target speeds, and also finer resolution.

The here presented results proof the suitability of the fiber-coupled COFDR system to determine Doppler-corrected position and velocity of very small targets over short distances with fine resolution and high accuracy for high-precision localization and tracking applications.

## Acknowledgements

E. Renner is part of the Max Planck School of Photonics supported by BMBF, Max Planck Society, and Fraunhofer Society.